\documentclass[
 reprint,
 superscriptaddress,
 nofootinbib,
 amsmath,amssymb,
 aps,
 prl,
]{revtex4-2}

\usepackage{graphicx}% Include figure files
\usepackage[usenames,dvipsnames]{color}
\usepackage{multirow}
\usepackage[
        range-phrase=--,
        range-units=single,
        retain-explicit-plus=true,
        retain-unity-mantissa=false,
    ]{siunitx}

\usepackage[dvipsnames,x11names]{xcolor}
\usepackage[
        colorlinks=true,
        citecolor=RoyalBlue,
        linkcolor=BrickRed,
        urlcolor=DarkGreen,
    ]{hyperref}
\usepackage{url}

\newcolumntype{P}[1]{>{\centering\arraybackslash}p{#1}}

\definecolor{spring}{rgb}{0.7,0.9,0.7}
\definecolor{brick}{rgb}{0.7,0.2,0.1}
\definecolor{redHL}{rgb}{1.0,0.7,0.7}
\definecolor{blueHL}{rgb}{0.7,0.7,1.0}
\definecolor{ElectricBlue}{rgb}{0.49, 0.976, 1.0}
\definecolor{blueT}{rgb}{0.039, 0.729, 0.71}
\definecolor{DarkGreen}{rgb}{0, 0, 0}
\definecolor{blue}{rgb}{0.066, 0.310, 0.984}
\definecolor{black}{rgb}{0, 0, 0}
\usepackage[capitalise]{cleveref}

\newcommand{\hh}[1]{{\color{black}#1}}

\newcommand{\rme}{\mathrm{e}}

\newcommand{\rmd}{\mathrm{d}}

\newcommand{\mysec}[1]{\textit{#1~-~}}

\begin{document}

\preprint{APS/123-QE}

\title{Point Absorber Limits to Future Gravitational-Wave Detectors}

% \newgeometry{left=1.8cm, right=1.8cm, top=1.8cm, bottom=2.6cm}

\author{Wenxuan Jia}
\email{wenxuanj@mit.edu}
\affiliation{LIGO, Massachusetts Institute of Technology, Cambridge, MA 02139, USA}
\author{Hiroaki Yamamoto}
\affiliation{LIGO, California Institute of Technology, Pasadena, CA 91125, USA}
\author{Kevin Kuns}
\affiliation{LIGO, Massachusetts Institute of Technology, Cambridge, MA 02139, USA}
\author{Anamaria Effler}
\affiliation{LIGO Livingston Observatory, Livingston, LA 70754, USA}
% \author{\\Aidan F. Brooks}
% \affiliation{\LigoCaltech}
\author{Matthew Evans}
\affiliation{LIGO, Massachusetts Institute of Technology, Cambridge, MA 02139, USA}
\author{Peter Fritschel}
\affiliation{LIGO, Massachusetts Institute of Technology, Cambridge, MA 02139, USA}

% 
% LSC list
%
\author{R.~Abbott}
\affiliation{LIGO, California Institute of Technology, Pasadena, CA 91125, USA}
\author{C.~Adams}
\affiliation{LIGO Livingston Observatory, Livingston, LA 70754, USA}
\author{R.~X.~Adhikari}
\affiliation{LIGO, California Institute of Technology, Pasadena, CA 91125, USA}
\author{A.~Ananyeva}
\affiliation{LIGO, California Institute of Technology, Pasadena, CA 91125, USA}
% ...

% \collaboration{LSC Instrument List}%\noaffiliation

\begin{abstract}
High-quality optical resonant cavities require low optical loss, typically on the scale of parts per million. However, unintended micron-scale contaminants on the resonator mirrors that absorb the light circulating in the cavity can \hh{deform the surface thermoelastically}, and thus \hh{increase losses} by scattering light out of the resonant mode. The point absorber effect is a limiting factor in some high-power cavity experiments, for example, the Advanced LIGO gravitational wave detector. In this Letter, we present a \hh{general} approach to the point absorber effect from first principles and simulate its contribution to the increased scattering. The \hh{achievable circulating power} in current and future gravitational-wave detectors is calculated statistically given different point absorber configurations. Our formulation is further confirmed experimentally in comparison with the scattered power in the arm cavity of Advanced LIGO measured by in-situ photodiodes. The understanding presented here provides an important tool in the global effort to design future gravitational wave detectors that support high optical power, and thus reduce quantum noise.

\end{abstract}

\maketitle

\mysec{Introduction}
A wide variety of precision optical experiments rely on resonant optical cavities to enable precise measurements of space, time, and fundamental physics.
These experiments often require high optical intensity incidents on the mirrors of the cavity to boost the signal-to-noise ratio.
However, unintended defects may be deposited on the reflective surface of the mirror during the coating process or exposure to a dusty environment \cite{Aidan}. These localized defects, known as ``point absorbers'', absorb optical power and cause undesired thermal effects on the optics under irradiation, especially in cavities containing high circulating power. The point absorber becomes a limiting factor in various precision measurement experiments that require a high-finesse cavity with low round-trip loss, such as cavity QED \cite{cavityQED}, axion detection \cite{axion}, qubit experiments \cite{qubit}, and gravitational-wave detectors \cite{aLIGO, aVirgo, KAGRA}. It is thus necessary to develop a quantitative understanding of the point absorber effect in high-power optical cavities.

With a \SI{4}{\km} long baseline and a circulating power of more than \SI{200}{\kilo\W}, the arm cavity of Advanced Laser Interferometer Gravitational-Wave Observatory (aLIGO) serves as a good example of the point absorber effect \cite{Aaron}.
aLIGO is a dual-recycled Fabry--P\'{e}rot Michelson interferometer designed to measure tiny perturbations of spacetime with unprecedented precision \cite{Peter}.
One of the fundamental noises that limit aLIGO's performance is quantum shot noise, which can be reduced either by increasing the arm power or by manipulating the quantum states of light through squeezing \cite{Maggie}.
However, arm power can be limited by
point absorbers (studied here),
other thermal distortions \cite{thermalLensing},
and a variety of instabilities
\cite{PI, PI2, AI1, AI2}.
The arm power during the third observing run was limited to one-third of the designed value of \SI{750}{\kilo\W}, mainly due to point absorbers on the mirror \cite{Aaron, Peter} that scatter light out of the fundamental cavity mode.

Point absorbers were known to exist since the first observing run.
Many analyses have been carried out to understand how they deform the optics and scatter light out of the cavity \cite{Aidan, PA2, Evan}. In this paper, we provide a more general approach from first principles. The traditional formalism is extended to include arbitrary heating functions with any nonlinear boundary conditions, such as Stefan-Boltzmann law. With the correction from nonlinear boundary condition, we can make more accurate statistical estimations of the arm power for the next planned upgrade of aLIGO (known as ``A+'') and the next generation of gravitational-wave detectors with a variety of potential point absorber configurations. 

We start by calculating the differential temperature profile from single point absorber heating with proper boundary conditions. Then the thermoelastic deformation of the mirror is derived using thermoelasticity equation.
Next, this deformation is incorporated in an FFT-based simulation to obtain the field in the arm cavity, which is used to calculate its round-trip loss and achievable power.  In addition, we simulate the low-angle scattered light intensity and compare this with in-situ measurements. Our results reveal a good match between these measurements and simulation, thus confirming our understanding of point absorbers.

%%%%%%%%%%%%%%%%%%%%%%%%%%%%%
\mysec{Theoretical modeling}
Point absorbers degrade the performance of high-power optical resonators by absorbing laser power, which thermally distorts the mirror surface and thereby scatters light out of the resonant mode of the cavity.
The analytical solution of the differential temperature under a general boundary condition is derived first.

Consider a cylindrical optic with radius $a$ and thickness $h$.
Choose cylindrical coordinates at the center of the mirror with the $z$ direction pointing into the cavity.
One point absorber is put at the center of the high-reflective (HR) side for cylindrical symmetry. When the cavity is held on resonance, the system is static, and the heat equation reduces to the Laplace equation
\begin{equation}
    \nabla^2 T = 0
\end{equation}
where $T(r,z)$ is the temperature departure from the ambient temperature $T_{\infty}$. The boundary conditions include the intensity of a heating source $I(r)$ on the HR surface due to the point absorber:
\begin{equation} \label{eq: frontSurfaceBC}
    -K\frac{\partial T}{\partial z} \bigg|_{z = \tfrac{h}{2}} = -I(r) +  g(T)\bigg|_{z = \tfrac{h}{2}}
\end{equation}
where $K$ is the thermal conductivity, 
$g(T)$ is the thermal flux of blackbody radiation
\begin{equation}
    g(T) = \epsilon \sigma \left[ \left(T_{\infty}+T(r,z) \right)^4 - T_{\infty}^4 \right],
\end{equation}
$\epsilon$ is the thermal emissivity/absorptivity (assumed to be unity throughout this paper), and $\sigma$ is the Stefan-Boltzmann constant.

A semi-infinite assumption can be made by treating the optic as a semi-infinite solid \cite{HelloT,Lu}. We present a general way of solving the Laplace equation under either linearized or nonlinear boundary conditions.
Taking the zeroth-order Hankel transform $\mathcal{H}_0$ of the Laplace equation in cylindrical coordinates with respect to $r$ \cite{Hankel}, we get
\begin{equation} \label{eq: PDE_T(k)}
    (-k^2  + \partial_z^2)~\tilde{T}(k, z) = 0
\end{equation}
The angular dependence is dropped due to cylindrical symmetry. The solution of \cref{eq: PDE_T(k)} is the sum of growth modes $A(k)\rme^{kz}$ and decay modes $B(k)\rme^{-kz}$, where the latter vanishes by the semi-infinite boundary conditions $T(r \rightarrow \infty,z) = T(r,z \rightarrow -\infty) = 0$.

Let $I(r)$ be the heating function from a point absorber; here we use a Gaussian profile $I(r) = \epsilon I_{\text{b}} \rme^{-r^2/w^2}$ with absorber radius $w$ and irradiation intensity $I_{\text{b}}$ at the absorber center. Let $T_{\text{HR}}(r) = T(r, z=h/2)$ be the temperature profile at the HR surface. The Hankel transform of the boundary condition at the HR surface \cref{eq: frontSurfaceBC} gives
\begin{equation} \label{eq: T(k,z)}
    \tilde{T}(k, z) = -\frac{\rme^{k(z-h/2)}}{Kk} \mathcal{H}_0\left[-I(r) +  g(T_{\text{HR}}(r)) \right]
\end{equation}
Therefore, $T_{\text{HR}}(r)$ is found by taking the inverse transform of $\tilde{T}\left(k,z=h/2\right)$:
\begin{equation} \label{eq: sol}
\begin{aligned}
    T_{\text{HR}}(r) 
    &= \frac{2}{\pi K} \int_0^r \rmd r' r' \left[I(r') - g\left(T_{\text{HR}}(r')\right)\right] \frac{1}{r} \mathcal{K}\left(\frac{r'^2}{r^2} \right)\\
    & \ \ \ \ +\frac{2}{\pi K} \int_r^{\infty} \rmd r' \left[I(r') - g\left(T_{\text{HR}}(r')\right)\right] \mathcal{K}\left(\frac{r^2}{r'^2} \right)\\
    \end{aligned}
\end{equation}
where $\mathcal{K}$ is the complete elliptic integral of the first kind. \cref{eq: sol} is a nonlinear integral equation with no closed-form solution, but an approximate solution can be found by either linearizing the boundary function $g(T)$ or using successive approximation.

%%%%%%%%%%%%%%%%%%%%%%%%%

The linearized boundary solution has been given in \cite{Lu}. The heat from a small point absorber is primarily dissipated by conduction. This estimate breaks down for a large absorber of radius $(K/\epsilon I_\text{b})(I_\text{b}/2\sigma)^{1/4}$ which is ${\sim}\SI{100}{\um}$ in an aLIGO arm cavity \cite{Evan}. The correction of nonlinear radiation will matter if the radiative contribution becomes significant. This motivates us to find a solution to the general boundary condition. 
%%%%%%%%%%%%%%%%%%%%%%%%%
\begin{figure}[t]
\begin{center}
	\includegraphics[width=0.8\linewidth]{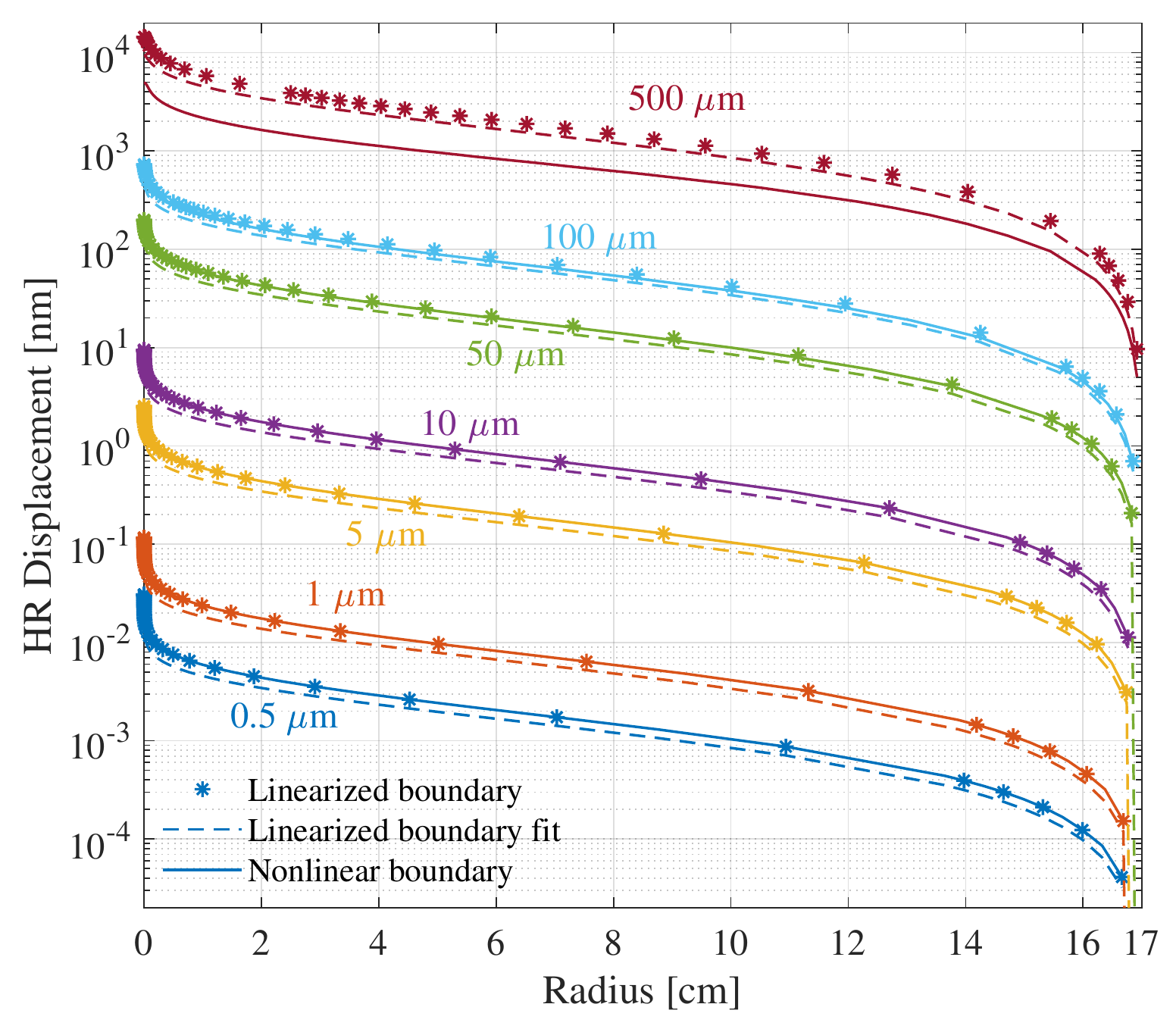}
	\caption{Thermoelastic displacements on the HR surface by various point absorber radii (labeled near each curve). The edge of the 17-cm radius optic has zero deformation. The incident intensity on the centered absorber is \SI{4.1e7}{W/m^2}, equivalent to the center intensity of the \SI{240}{\kilo\W} beam on the mirror of aLIGO arm cavity. Analytic fits to the linearized boundary solution (\cref{eq: h_analytical}) are also shown. \label{fig: h}}
\end{center}
\end{figure}

The nonlinear integral equation Eq.~(\ref{eq: sol}) can be solved by successive approximation with feedback. We start the zeroth iteration with an initial guess $T_0(r)$. The real solution is denoted as $T_S(r)$, and the zeroth error function is $\varepsilon_0(r) = T_0(r) - T_S(r)$. Plugging this into \cref{eq: sol} and keeping the first non-trivial order of the error:
\begin{equation}
    g_0(r) = g(T_0(r)) = g_S(r) + 4\epsilon \sigma (T_{\infty}+T_0)^3\varepsilon_0 + \mathcal{O}(\varepsilon^2)
\end{equation}

%%%%%%%%%%%%%%%%%%%%%%%%%
\begin{figure*}[t]
\begin{center}
	\includegraphics[width=0.75\linewidth]{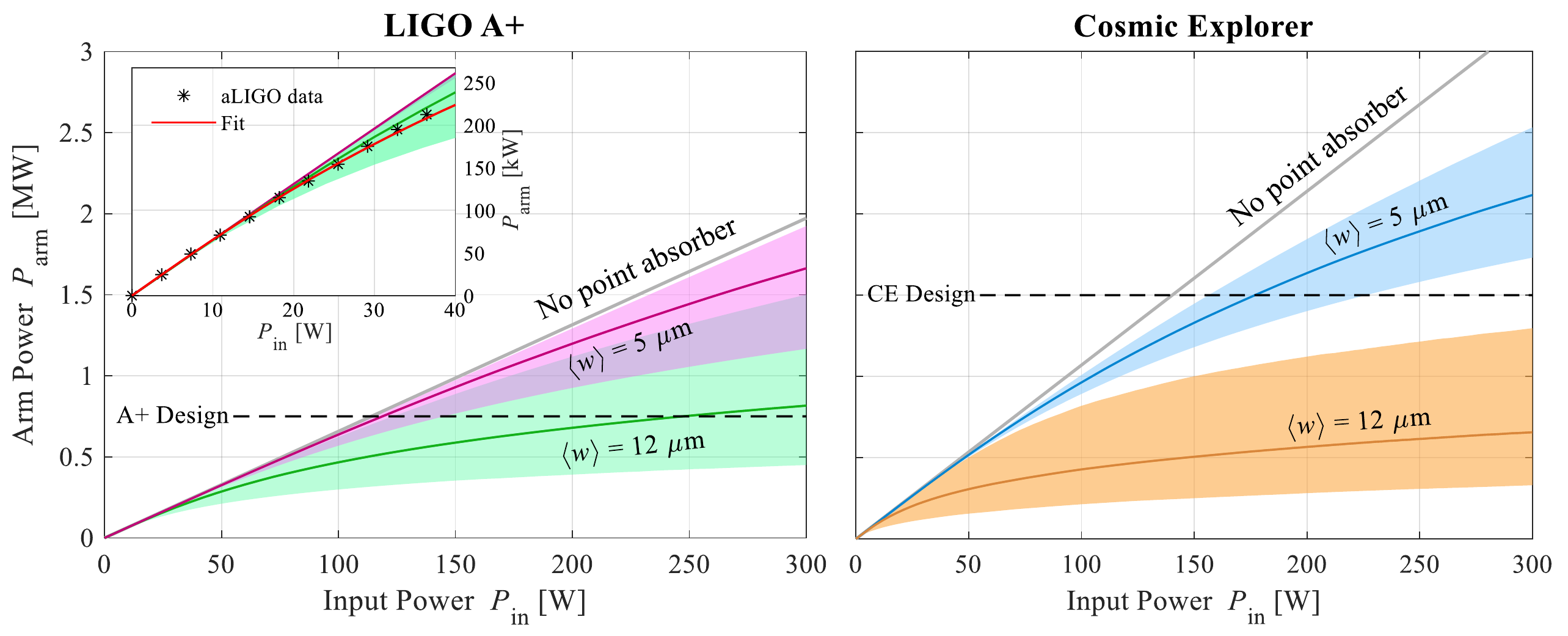}
	\caption{Circulating power in the arm cavity versus input power for two different detectors and mean radii of point absorbers (optimistic \SI{5}{\um} and pessimistic \SI{12}{\um}). The solid line is the median with shadings corresponding to the 16th and 84th percentile. The gray lines (no absorber case) increases linearly with the initial slopes set by the round-trip loss of the cold cavity (\cref{table}), and the designed power is \SI{750}{\kilo\W} for A+ and \SI{1.5}{\mega\W} for CE. In the absence of point absorbers, the required input power is \SI{120}{\W} for A+ and \SI{140}{\W} for CE. In the zoomed-in graph, the data points collected from LIGO Livingston Observatory throughout Observing Run 3b are fit to obtain the radii of point absorbers. It is statistically more confident to achieve the designed power with $\langle w \rangle = \SI{5}{\um}$.}
	\label{fig: PPplot}
\end{center}
\end{figure*}

Assuming the error has weak variation over radius: $\varepsilon_0(r) \approx \varepsilon_0$, we have
\begin{equation}
    T(T_0(r)) = T_S(r) - \varepsilon_0 C_0(r) = T_S(r) - (T_0(r)-T_S(r)) C_0(r) 
\end{equation}
where
\begin{equation}
\begin{split}
    C_0(r) 
    &= \frac{8\epsilon \sigma }{\pi K} \left[ \int_0^r \rmd r' \frac{r'}{r} (T_{\infty}+T_0)^3 \mathcal{K}\left(\frac{r'^2}{r^2} \right) \right. \\
    & \ \ \ \ \ + \left. \int_r^{\infty} \rmd r' (T_{\infty}+T_0)^3 \mathcal{K}\left(\frac{r^2}{r'^2} \right) \right]
\end{split}
\end{equation}
We can then iterate the temperature profile
\begin{equation}
    T_{i+1} = \frac{T(T_i) + C_{i} T_{i}}{1+C_{i}}
\end{equation}
until we reach convergence at $T(T_i) = T_i$ and $T_{i+1} = T_i$ \cite{nlinIntEq1, nlinIntEq2}.

The displacement vector field of the optic can be found given the temperature solution $T(r,z)$. We follow Hello and Vinet's formalism but apply it to our solution \cite{Helloh} (see the supplemental material for detailed derivations). \cref{fig: h} shows the resulting displacement of the HR surface. For small point absorbers, the differential temperature is relatively low, and the radiative correction is negligible; the linearized boundary solution is accurate. However, the correction becomes significant for absorber with radius larger than \SI{100}{\um}, up to a factor of three in the \SI{500}{\um} case. A typical point absorber with a few tens of microns in radius can cause surface deformation on the scale of several tens of nm in height and a few cm in size. 

An analytical fit to displacement of the linearized solution is given by
\begin{equation} \label{eq: h_analytical}
    h(r) \approx 0.12 \left( \frac{3\lambda+2\mu}{\lambda+\mu}\right) \frac{ \epsilon I_{\text{b}} w^2 \alpha}{K} \ \ln\left(\frac{a^2}{r^2 \left(1-\frac{w^2}{a^2}\right) +w^2} \right)
\end{equation}
where $\mu$ is the first Lam\'{e} coefficient, $\lambda$ is the second Lam\'{e} coefficient, and $\alpha$ is the thermal expansion coefficient. Note that \cref{eq: h_analytical} breaks down at high absorbed power, as shown in \cref{fig: h}. With the deformation known, we can superpose it onto the mirror phase map data and simulate fields in a static cavity.

\mysec{Implications for Gravitational Wave Detectors}
Advanced gravitational-wave detectors are Michelson interferometers using Fabry–P\'{e}rot cavities as arms to increase optical power and thus the signal produced by gravitational-wave strain. The arm power is further increased by the addition of a mirror at the symmetric port of the interferometer to form a power-recycling cavity \cite{aLIGO}. However, the power buildup can be degraded by the point absorber effect as follows.

Without any thermoelastic deformation, the round trip loss in the cavity is constant, and the arm power is linearly proportional to the input power with the slope set by the round-trip loss of the cold cavity (gray lines in \cref{fig: PPplot}). However, the thermoelastic deformation from the point absorbers contributes to the optical loss by scattering light out of the fundamental cavity mode. Thus, an increase in arm power leads to an increase in the optical loss of the arm, which decreases the optical gain of the power recycling cavity \cite{Aidan}. As a result, for sufficiently high power levels the arm power saturates and becomes largely independent of the input power.

Understanding the limitations of point absorbers on the achievable arm power is important in planning future detectors, for example, the next-generation gravitational-wave detector Cosmic Explorer (CE) \cite{CE, CE2}. CE will achieve a factor of ten increase in sensitivity relative to A+ by scaling up the A+ design to use \SI{40}{\km} long arm cavities and increasing the arm power by a factor of two. The key parameters of the coupled arm cavities of both detectors are summarized in \cref{table}.

To investigate the achievable arm power in CE and the upcoming A+ observing runs, we conducted a statistical analysis of round-trip loss by calculating fields under a thousand point absorber maps generated on the arm cavity mirrors. For each map, the absorber locations are uniformly distributed; radii are governed by a Rayleigh distribution, and number are governed by a Poisson distribution with mean number density one per \SI{60}{cm^2}, characteristic of coated aLIGO mirrors.
We investigate the cases of mean absorber radius $\langle w\rangle = \SI{5}{\um}$ (optimistic) and larger absorbers with $\langle w\rangle = \SI{12}{\um}$ (pessimistic). The FFT-based simulation package Stationary Interferometer Simulation (SIS) \cite{SIS} is used to calculate the field amplitudes in the cavity given these point absorber maps. The round-trip loss for each map is calculated at several arm powers from which the power recycling gain is computed. The recycling gain is then converted to the input power required to reach a given arm power.

%%%%%%%%%%%%%%%%%%%%%%%%%%%
\begin{table}[b] 
\caption{Parameters of Y-arm cavity of LIGO Livingston Observatory and the proposed Cosmic Explorer.} \label{table}
\begin{ruledtabular}
\begin{tabular}{p{0.5\linewidth} P{0.25\linewidth} P{0.25\linewidth}}
{Parameter} &  {aLIGO} & {CE} \\
     \colrule
     Designed arm power & \SI{750}{\kW} & \SI{1.5}{\MW} \\
     
     {Optical gain of:} &   \\
     {\hspace{0.5cm} Power recycling cavity}  & 40 & 76 \\
     {\hspace{0.5cm} Arm cavity}  & 270 & 280\\
     
    %  Arm cavity finesse & 420 & 440\\
     
     {Round trip loss of:} &   \\
     {\hspace{0.5cm} Power recycling cavity}  & \SI{500}{ppm} & \SI{500}{ppm} \\
     {\hspace{0.5cm} Cold arm (no absorber)}  & \SI{66}{ppm} & \SI{40}{ppm} \\
     
    %  {Radius of curvature of:} &   \\
    %  {\hspace{0.5cm} ITM}  & \SI{1941}{m} & \SI{30}{km} \\
    %  {\hspace{0.5cm} ETM }  & \SI{2245}{m} & \SI{30}{km} \\
     
     Cavity length & \SI{3995}{m} & \SI{40}{km} \\
     
     Mirror &   \\
     {\hspace{0.5cm}} Aperture & \SI{34}{cm}& \SI{70}{cm} \\
     {\hspace{0.5cm}} Material & Fused Silica& Fused Silica \\
     {\hspace{0.5cm}} Temperature & \SI{290}{K}& \SI{290}{K} \\

     Beam radius on: & \\
     \hspace{0.5cm} Input mirror & \SI{5.2}{cm} & \SI{12}{cm} \\
     \hspace{0.5cm} End mirror & \SI{6.1}{cm} & \SI{12}{cm} \\

\end{tabular}
\end{ruledtabular}
\end{table}

%%%%%%%%%%%%%%%%%%%%%%%%%%%%%

\cref{fig: PPplot} shows the results for these two cases for both the A+ and CE arm cavities. The medians are shown as solid lines and the shadings correspond to the 16th and 84th percentile. The arm power saturation is evident in the $\langle w\rangle = \SI{12}{\um}$ case and, while it may be possible for A+ to reach its \SI{750}{\kilo\W} design arm power, it is unlikely that CE would ever reach its design of \SI{1.5}{\mega\W} with absorbers of this size. On the other hand, our analysis suggests that point absorbers with $\langle w \rangle = \SI{5}{\um}$ pose little risk of damaging the A+ arm power, but it requires on average 30\% more input power for CE to achieve the designed goal. In both cases, the point absorbers limit the arm power of CE more significantly than that of A+.

%%%%%%%%%%%%%%%%%%%%%%%%%
\begin{figure}[t!]
\begin{center}
		\includegraphics[width=\linewidth]{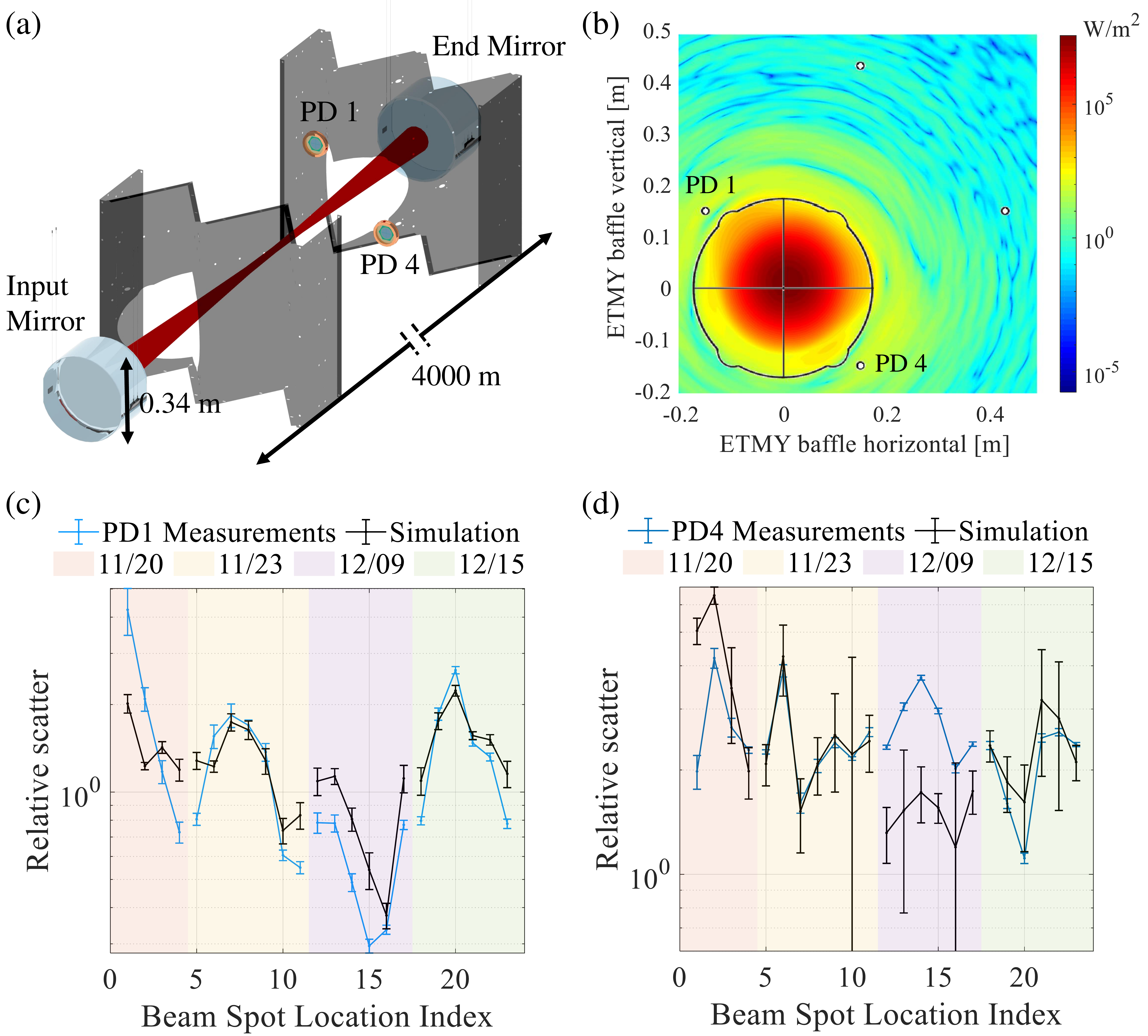}
        \caption{(a) Schematics of the Y arm cavity of LIGO Livingston Observatory with photodiodes (PD 1 and 4) marked. (b) Intensity distribution of the field incident on the end mirror baffle with a through hole at origin. (c-d) Experimental measurement (with $5\times$error bar) of nomalized scatter power landing on PD 1 and 4 versus FFT simulation with point absorber formulation incorporated. The error bar of simulation is due to the \SI{3}{\mm} uncertainty of beam position. The data is taken at 23 beam spot locations on the end mirror at four different days. The relative scatter of clean optics without any absorber is roughly an order of magnitude lower than the plotted simulation curve (not shown).}
	\label{fig: ACBPD}
\end{center}
\end{figure}

This statistical model is consistent with measured arm powers in the LIGO Livingston observatory during Observing Run 3, which deviate from linear growth at high power due to the point absorber effect. This data, shown in the inset graph of \cref{fig: PPplot}, is fit to yield a \SI{12.6}{\um} radius absorber and \SI{66}{ppm} round trip loss of the cold cavity. The thermal absorptivity is taken as unity to break its degeneracy with the radius of the point absorber (\cref{eq: h_analytical}). The data sits in the predicted region of the pessimistic case. These results are also consistent with measurements of the total absorbed power of the point absorber \cite{alog_aidan_ETMY_HWS}. 

%%%%%%%%%%%%%%%%%%%%%%%%%

\mysec{Scatter Magnitude}
Knowing the absorber radii, beam position, and cavity parameters, we can calculate the scattered fields through FFT simulation and compare the theoretical modeling with measurements. Inside the arm cavity, there are four silicon photodiodes (PDs) mounted on each of the baffles installed in front of the test mass optics to block and monitor scattered light. As the power in the interferometer increases, the absorbers cause thermoelastic aberration of the HR surface of the test mass, which in turn results in an increased scattering. PD 1 and 4 facing the cavity sample the Airy patterns of scattered light, as shown in \cref{fig: ACBPD}(a-b). 

There were roughly a dozen point absorbers scattered around the surface of the end mirror, including one dominant absorber with the largest size near the center of the mirror. After the Observing Run 3, we moved the beam spot at 23 locations on the end mirror to change the intensity incident on the absorbers while fixing the beam spot on the input mirror. Simulations of each of these 23 alignments reveals that this large and centrally located absorber dominates the optical scatter
(\cref{fig: PPplot}). 
The FFT results are shown in \cref{fig: ACBPD}(c-d) for each of the 23 spot locations. We moved the beam to the same location repeatedly at indices 12, 17, 18, and 23. It is seen that the measurements at these indices are equal, indicating that our measurements are reproducible over a week. The simulation is capable of predicting the magnitude and variation of the low-angle scatter, even though the field amplitude shows a great amount of structure along the radial distance from the beam center. The consistency between data and simulation lends further credibility to our modeling and improves our understanding of the point absorber effect. Without the scattering due to point absorbers, the simulated relative scatter magnitude is roughly a factor of ten lower, and the simulated variations show little coherence with the PD measurements.

%%%%%%%%%%%%%%%%%%%%%%%%%%%%%
\mysec{Conclusion}
In summary, we carried out an analytical approach to the point absorber problem in a high-power resonant cavity. We propose an analytical solution to the thermoelastic deformation of the optics with arbitrary point absorber heating function and boundary conditions. Both temperature and displacement profiles are derived and incorporated in the state-of-the-art FFT-based optical simulation. With a more advanced and accurate understanding of the point absorber effect, we make a statistical prediction of arm power in current and future gravitational-wave detectors for different mean radii of point absorbers. Our analysis of resonant field power in the cavity suggests that point absorbers of mean \SI{5}{\um} radii will not prevent future gravitational-wave detectors from achieving their design sensitivity. Active research is being carried out to mitigate both the size and number of point absorbers on future optics. Finally, our formulation shows a strong coherence with data when compared with in-situ measurements of scattered light, thus confirming our model. 

Future analyses on the distortion of phase and mode-shape of the fields from point absorbers are needed to estimate the degradation on the Michelson contrast, which impacts the signal-to-noise ratio and thus the sensitivity of the gravitational wave detectors.

\vspace{0.2cm}
%%%%%%%%%%%%%%%%%%%%%%%%%%%%%
\mysec{Acknowledgements}
The author acknowledges the support of MathWorks Science Fellowship and Sloan Foundation,
 and thanks The MathWorks Inc. for its generous computing support.
aLIGO was constructed by the California Institute of
Technology and Massachusetts Institute of Technology with funding from the NSF and operates under Cooperative Agreement No. PHY-1764464. aLIGO was built under Award No. PHY-0823459.

\bibliography{point_absorber}% Produces the bibliography via BibTeX.

\end{document}